# Temperature dependent Casimir forces: recurring subtleties.

L.R. Fisher* & B.W. Ninham**

*School of Physics, University of Bristol, UK (Corresponding author: [len.fisher@bristol.ac.uk](len.fisher@bristol.ac.uk))

**Emeritus Professor, Research Scholl of Applied Mathematics, Australian National University

The Casimir[1] force between two ideal conducting surfaces is a special (zero temperature) limit of a more general theory due to Lifshitz[2]. Extensions of this theory of van der Waals forces between dielectric plates to include an intermediate medium were derived from quantum field theory by Dzyaloshinski *et al*[3]. The mystique that attended this apparently complete quantum mechanical solution of the many-body interaction problem was tempered by the later realization that it can also be derived semi-classically from Maxwell's equations plus Planck quantization of light[4]. The temperature dependent theory includes correlations in coupled quantum and classical fluctuation modes for conducting, dielectric, and magnetic media. How this happens in a vacuum is explained in ref. 5. If the surfaces are at different temperatures, it has been postulated (e.g.[6]) that these modes might act as a coupling spring, transferring thermal energy from the hotter to the colder. Fong *et al*[7] have used a cleverly designed nanomechanical system to test this effect for forces between gold-plated surfaces. They compare their data with Casimir's original theoretical expression. This expression, however, is just the leading term in an expansion, valid, if at all, only at zero temperature. It is the finite temperature expression[8] with which the measurements should be compared. Another issue is that real gold surfaces require a correction factor of up to 25% compared to ideal conducting surfaces[9,10]. This correction factor is both distance- and temperature-dependent. We give numerical values for these corrections. It appears that they may not affect the basic conclusions of Fong *et al*, but the take-home message is that care is needed in the interpretation of extensions of Casimir (Lifshitz) effects, which are increasingly emerging across a wide range of scientific problems.

The temperature-dependent expression for perfectly conducting surfaces takes two limiting forms (Eqs. 33 and 35 of ref. 8, corresponding to dotted lines in Fig.1 of ref. 10). The high temperature/large distance approximation ($2kTl/\hbar c \gg 1$) is

$$E(l,T) = \frac{kT}{8\pi l^2}\left[\zeta(3)+\left(\frac{8\pi kTl}{\hbar c}+2\right)e^{-4\pi kTl/\hbar c}+O(e^{-8\pi kTl/\hbar c})\right]$$

While the low temperature/small distance approximation ($2kTl/\hbar c \ll 1$) is

$$E(l,T) = -\frac{\pi^2 \hbar c}{720 l^3} - \frac{\zeta(3)(kT)^3}{2\pi(\hbar c)^2} + \frac{(kT)^4 \pi^2 l}{45(\hbar c)^3} - \frac{(kT)^2}{\hbar c l}\left(1+\frac{kTl}{\pi\hbar c}\right)e^{(-\pi\hbar c/kTl)} + O(e^{-(2\pi\hbar c/kTl)})$$

where $l$ is the distance between the surfaces, $E(l,T)$ is the free energy per unit area, and all of the other symbols have their usual meaning.

Under the conditions used by Fong *et al*, $2kTl/\hbar c$ ranges from 0.08 (at 300nm) to 0.22 (at 800nm), so the small distance approximation is (barely) justified. The first term in this equation corresponds to the Casimir expression as used by them. The second term is a chemical potential contribution due to virtual electron-positron pairs in the vacuum, and the third term arises from black body radiation between the surfaces. The ratio of the second term to the first increases rapidly with distance. Fortunately, at Fong *et al*'s experimental temperature of 300K, and their greatest experimental distance of 800nm, the ratio is only 1.6% (the ratios for the subsequent terms are considerably smaller), so that their use of Casimir's original equation turns out to be numerically reasonable for ideal surfaces.

Gold surfaces, however, are not ideal, and a correction factor needs to be applied to the Casimir-Lifshitz free energy as calculated from the first term for the low temperature/small distance approximation (Casimir's original expression, as used by Fong *et al*). The correction factors are (solid curves in Fig.1 of ref. 10):

| Separation (nm) | Correction at 0K | Correction at 300K |
|---|---|---|
| 300 | x0.74 | x0.69 |
| 400 | x0.79 | x0.73 |
| 500 | x0.82 | x0.74 |
| 600 | x0.85 | x0.75 |
| 700 | x0.87 | x0.75 |
| 800 | x0.88 | x0.75 |

So far as we can tell, these corrections do not invalidate the conclusions of Fong *et al*. The point is important, however, because the real world involves temperature-dependent Casimir-Lifshitz forces, which are significant in areas that range from the design of nanoelectromechanical systems (NEMS)[11] to the search for dark matter[12]. The correct interpretation of the forces involved requires correct theory as its foundation. Simple approximations may be sufficient, but care is needed in their use. One example is the so-called "retarded ground state – excited state interaction" problem, where simplified theory even gets the zero temperature limit badly wrong[13] (worse, the classical interpretation of "retardation" is physically incorrect[14]). The connection between Casimir-Lifshitz forces and weak nucleon interactions mediated by mesons[10] also requires the temperature-dependent form of the equation.

We note in particular that Lifshitz temperature-dependent forces are intimately related to a realistic treatment of quantum entanglement[15,16], usually analyzed only at zero temperature. Temperature dependence is also essential when it comes to fundamental issues such as connections between quantized electromagnetic theory and weak interactions[16] and Bell's inequality[17], as well as the practical issues involved in quantum computing and other quantum-based technologies.

The interpretation of temperature-dependent Casimir-Lifshitz forces involves many subtleties that are not always recognized in the literature. Simplifications and misinterpretations are endemic. In offering these comments on one specific and otherwise excellent paper, our aim is to draw attention to this general problem rather than to criticize this specific work.